\documentclass[letterpaper]{article} % DO NOT CHANGE THIS
\usepackage{aaai22}  % DO NOT CHANGE THIS
\usepackage{times}  % DO NOT CHANGE THIS
\usepackage{helvet}  % DO NOT CHANGE THIS
\usepackage{courier}  % DO NOT CHANGE THIS
\usepackage[hyphens]{url}  % DO NOT CHANGE THIS
\usepackage{graphicx} % DO NOT CHANGE THIS
\usepackage{natbib}  % DO NOT CHANGE THIS AND DO NOT ADD ANY OPTIONS TO IT
\usepackage{caption} % DO NOT CHANGE THIS AND DO NOT ADD ANY OPTIONS TO IT
\DeclareCaptionStyle{ruled}{labelfont=normalfont,labelsep=colon,strut=off} % DO NOT CHANGE THIS
\usepackage{algorithm2e}
\usepackage{multirow,color}
\usepackage{amsmath,amsthm,amssymb}
\usepackage{titlesec,adjustbox}
\usepackage{wrapfig, dsfont, cite}

\usepackage{newfloat}
\usepackage{listings}
\DeclareMathOperator*{\argmin}{arg\,min}
\newtheorem{defn}{Definition}
\newtheorem{thrm}{Theorem}

\SetKwComment{Comment}{/* }{ */}
\RestyleAlgo{ruled}

\newcommand\scalemath[2]{\scalebox{#1}{\mbox{\ensuremath{\displaystyle #2}}}}

\title{Maximizing Social Welfare in Selfish Multi-Modal Routing using Strategic Information Design for Quantal Response Travelers}
\author{
    Sainath Sanga, \\
    Venkata Sriram Siddhardh Nadendla, \\
    Sajal K. Das
}
\affiliations{
}

\begin{document}

\maketitle

\begin{abstract}
Traditional selfish routing literature quantifies inefficiency in transportation systems with single-attribute costs using price-of-anarchy (PoA), and provides various technical approaches (e.g. marginal cost pricing) to improve PoA of the overall network. Unfortunately, practical transportation systems have dynamic, multi-attribute costs and the state-of-the-art technical approaches proposed in the literature are infeasible for practical deployment. In this paper, we offer a paradigm shift to selfish routing via characterizing idiosyncratic, multi-attribute costs at boundedly-rational travelers, as well as improving network efficiency using strategic information design. Specifically, we model the interaction between the system and travelers as a Stackelberg game, where travelers adopt multi-attribute logit responses. We model the strategic information design as an optimization problem, and develop a novel approximate algorithm to steer \textbf{Lo}git \textbf{R}esponse travelers towards social welfare using strategic \textbf{I}nformation design (in short, LoRI). We demonstrate the performance of LoRI on a Wheatstone network with multi-modal route choices at the travelers. In our simulation experiments, we find that LoRI outperforms SSSP in terms of system utility, especially when there is a motive mismatch between the two systems and improves social welfare. For instance, we find that LoRI persuades a traveler towards a socially optimal route for $66.66\%$ of the time on average, when compared to SSSP, when the system has 0.3 weight on carbon emissions. However, we also present a tradeoff between system performance and runtime in our simulation results.
% \textcolor{red}{SD: As mentioned before, is SSSP the state of the art method used in the problem context? I doubt it is! Otherwise the reviewers may criticize!}

\end{abstract}

\section{Introduction}
Smart navigation systems (e.g. GPS devices, navigation applications on mobile/smart devices) have transformed the transportation domain in terms of reducing cognitive overload in travelers. However, such technological advancements have had little impact on several fundamental issues such as mitigating congestion \cite{INRIX} and reducing carbon emissions \cite{energy}, which have only worsened over time. For instance, current state-of-the-art navigation systems employ traditional shortest paths algorithms, such as Dijkstra's algorithm \cite{lanning2014dijkstra}, Bellman–Ford or Warshall-Floyd, and $A^*$-algorithms \cite{10.5555/1614191} to recommend routes and mitigate travelers' cognitive overload. On the other hand, selfish travelers exhibit multi-attribute preferences, which are typically misaligned from system's interests. As a result, travelers often reject route recommendations that involve non-personal transport modalities, such as public transportation, ridesharing services and other micro-mobility services \cite{USCensus, mckenzie2015drives}. 
% For example, the pre-pandemic congestion cost in U.S. was valuated at about $\$1400$ per person in 2019 as per INRIX \cite{INRIX}. Although this cost varied voer time according to the economy and fuel prices, the consistent drive towards improving the economy and lowering fuel prices have led to persistent worsening of congestion costs in US. At the same time, there is a steady increase of transportation-related carbon emissions in the United States during 2012-2018, as per the U.S. Energy Information Administration. For instance, carbon emissions in the transportation sector alone amounted to around 1905 million metric tonnes of $CO_2$ in 2019, 25\% of which is due to personal cars \cite{energy}. 
Although unintentional, people have steered away from personal car usage during the ongoing COVID-19 pandemic in 2020 \cite{INRIX2020}, which have resulted in significant cost reductions in terms of congestion, carbon emissions as well as collisions. \textit {Our goal in this paper is to steer selfish travelers away from personal car usage (even under non-pandemic conditions), via offering them alternative routing choices in a persuasive manner.}
% U.S. drivers have been saving $\$980$ million in terms of congestion costs (which are valuated at $\$1374$ million and $\$394$ in 2019 and 2020 respectively). At the same time, collisions have also dropped by almost $30\%$ \cite{INRIX2020}.  

Selfish routing is a strategic framework where travelers employ their best-response routes selfishly according to their respective preferences to form an equilibrium. However, the central authority (e.g., a city transportation department) chooses a social-welfare objective that is not necessarily aligned with all travelers' interests. This leads to system inefficiency, which can be quantified by price-of-anarchy (PoA) \cite{roughgarden2002bad}. Several techniques have been proposed to drive PoA towards unity, which happens when the equilibrium outcome is optimal in terms of the system's objective. A seminal example is \emph{marginal cost pricing}, where selfish travelers are imposed taxes based on their marginal contribution to the system's objective \cite{ruggles1949recent}. Although the idea of marginal cost pricing has been floating around for several decades, the technique remains practically infeasible due to our inability to estimate marginal costs accurately. 

In \cite{sharon2019marginal}, the authors studied the effects of underestimating marginal costs on the optimality in terms of system objectives, and showed that taxing underestimating marginal costs produces an outcome that is at least as good as having no taxes. Although attempts have been made to implement such solutions by authoritarian regimes \cite{yang2020marginal}, the friction to adopt marginal cost pricing continues to persist due to various political reasons in democratic nations. Another powerful idea to influence traveler behavior is \emph{Stackelberg routing}, where a fraction of agents are routed centrally, while the remaining agents are allowed to choose their routes selfishly \cite{swamy2007effectiveness}. A similar routing algorithm is proposed In \cite{samal2018towards} based on multi-objective A$^*$ with a goal to design routes that decrease the overall network congestion. 

Meanwhile, information-revelation systems have also been proposed \cite{acemoglu2018informational, arnott1991does, mahmassani1991system}, where the traffic state is revealed to travelers as opposed to recommending routes. Although such systems do not mitigate cognitive-overload at the travelers, they have been found to generate a positive impact on traffic congestion and other global objectives even in non-strategic settings. However, these systems still suffer from poor persuasive ability, in terms of inducing behavior modification among travelers.
% Therefore, there is a need for a route system that considers each traveler's preferences over attributes to construct costs and personalised recommendations such that the traveler will not deviate from the recommended routes. Most of the route systems consider single-attribute cost functions at the traveler level to recommend routes leading to some problems, for example, consider a scenario where $1000$ travelers request a route between the same origin-destination pair at the same time $t$, these route systems might recommend the same route to all the $1000$ travelers leading to severe congestion. In other words, route recommendations system must consider the congestion across the entire network or at least congestion across the "zone" around the origin destination pair of a specific traveler.
A natural and effective solution is to design information strategically at the city transportation department, and present it to the travelers to steer their routing decisions towards socially optimal outcomes. 

Recently, strategic information design has been studied in the transportation domain when the network congestion state is uncertainly available at the travelers. For example, in \cite{das2017reducing}, the authors computed best-response signals under first-best, full information, public-signal and optimal information structure scenarios in the context of Wheatstone Network; they demonstrated that optimal information structures reveal only partial information revelation to mitigate network congestion. Similar results have been found in \cite{wu2019information} in the case of Pigou networks (graphs with parallel routes between a single-source and a single-destination) in the presence of state uncertainty on one of the routes. Optimal information structures have been found using Bayesian persuasion framework to reduce average traffic spillover on a specific route in a Pigou network. 

Despite the above development, existing works in strategic information design in transportation settings make several impractical assumptions. Since this is still a fledgling topic, almost all efforts assume that travelers are expected utility maximizers (EUM). However, there has been a strong evidence from real-world observations that travelers deviate from EUM behavior quite frequently. Such an effort was first made in \cite{nadendla2018effects}, which studied strategic information design in a single-sender, single-receiver setting when both are prospect-theoretic agents. Nevertheless, this framework is not applicable to transportation domain where there are multiple receivers. Another impractical assumption is the consideration of single-attribute costs and unimodal transportation networks, all of which are far from reality. Therefore, in this paper, we consider a more realistic transportation framework and develop a novel strategic information design framework as stated below.

First, we assume that the travelers' responses exhibit quantal response equilibrium (QRE), where deviations from EUM at each traveler are captured by the randomness within the stochastic utility maximization framework \cite{luce1959individual}. We model the strategic interaction with the system as a novel \emph{Stackelberg-QRE game}, where the system (leader) exhibits EUM behavior, while the travelers (followers) exhibit logit responses. Second, we assume that both the system and travelers exhibit non-identically weighted multi-attribute preferences. Specifically, we assume that the system's motive is to reduce both network congestion (in terms of travel time) and carbon emissions on the entire transportation network, whereas the traveler wishes to minimize travel time and/or carbon emissions along his/her personal route. 

Inspired from Bayesian persuasion  \cite{kamenica2019bayesian} as well as the method in \cite{bergemann2016bayes, mathevet2020information}, when there is a single sender and multiple receivers, we develop a novel, approximate strategic information design algorithm to steer \textbf{Lo}git \textbf{R}esponse travelers towards social welfare using strategic \textbf{I}nformation design (in short, LoRI). Our proposed algorithm LoRI uses the predictor-corrector method to find quantal responses at the travelers, and finds a locally-optimal state-information signal using interior-point algorithms that minimizes a non-convex system cost. Simulation results demonstrate that LoRI outperforms single source shortest path algorithms (e.g., Dijkstra's algorithm) 
% \textcolor{red}{SD: the same comment as in the Abstract} 
and improves social welfare in a Wheatstone network. We show that the system's cost reduces by $45\%$ when SSSP algorithm is designed with a misaligned objective function. 
% We also show that LoRI improves system cost by 45\%
% % \textcolor{red}{SD: mention by what percentage?} 
% even when a fraction of the travelers interact with LoRI, while other travelers use SSSP. 

\section{System Model and Problem Formulation}

\begin{figure}[!t]
\centering
\includegraphics[width=0.45\textwidth]{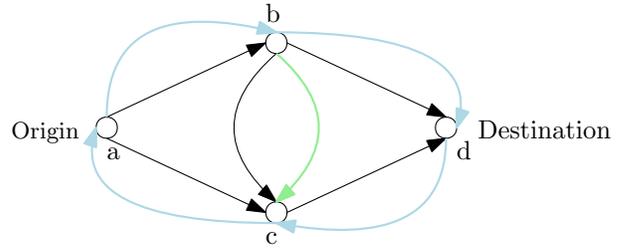}
\caption{An Example Multi-Modal Transportation Network}
\label{fig: Example Network}
\vspace{-3ex}
\end{figure}

\begin{figure*}[!t]
\centering
\includegraphics[width=\textwidth]{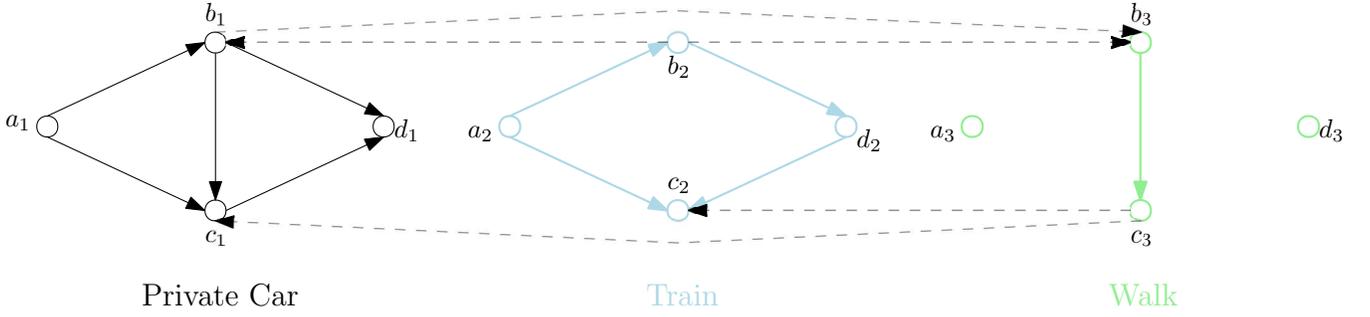}
\caption{Multi-layered expansion of the Multimodal Transportation Network shown in Figure \ref{fig: Example Network}}
\label{fig: Example Expanded Network}
\vspace{-2ex}
\end{figure*}

Let a multi-modal transportation network consisting of $\Lambda_t$ travelers at time $t$, be represented as a graph $\mathcal{G} = \{ \mathcal{V}, \mathcal{E} \}$, where $\mathcal{V} = \{0, 1, \cdots, N \}$ represents the set of physical locations (vertices), and $\mathcal{E}$ represents the transport interconnections (edges) between various locations in $\mathcal{V}$. Let $\mathcal{G}$ support a gamut of transport modalities 
%denoted by the set 
$\mathcal{M} = \{ 1, \cdots, M \}$. For the sake of convenience, we expand the network $\mathcal{G}$ into a multi-layered graph $\mathcal{G}_{exp.}$ using unimodal subgraphs $\{ \mathcal{G}_m \}_{m \in \mathcal{M}}$, and switch edge sets $\mathcal{E}_{i,j}$ which interconnect $i^{th}$ modality to $j^{th}$ modality within each vertex. 
% Consequently the edge set $\mathcal{E}$ is given by: 
% \vspace{-0.05in}
% \begin{equation}
%     \mathcal{E} = \displaystyle \Big( \bigcup_{i = 1}^M \mathcal{E}_i \Big) \cup \Big( \bigcup_{i \neq j} \mathcal{E}_{i,j} \Big). 
% \end{equation}
For example, consider a Wheatstone road network with four vertices and ten edges, as illustrated in Figure \ref{fig: Example Network}. Consider $M=3$ transport modalities on this network, and $\mathcal{M}$ =\{\emph{Private Car} (colored black), \emph{Metro Train} (colored blue) and \emph{Walking} (colored green)\}.
Using unimodal subgraphs and switch edges (depicted using dashed lines), we expand the example network into a multi-layered graph $\mathcal{G}_{exp.}$, as shown in Figure \ref{fig: Example Expanded Network}.  
% The vertices for mode privater car are $\mathcal{V}_{1} = \{a_1, b_1, c_1, d_1\}$, vertices for train mode are $\mathcal{V}_{2} = \{a_2, b_2, c_2, d_2\}$, vertices for walk modeing are $\mathcal{V}_{3} = \{a_3, b_3, c_3, d_3\}$. All the \textit{black} edges are of car mode, all the \textit{blue} edges are of train mode and all the \textit{green} edges are of walk mode. 
We model the network state as $s_t = \left\{ c_{e,t} \right\}_{e \in \mathcal{E}}$, where $c_{e,t}$ is the number of travelers on edge $e \in \mathcal{E}$ at time $t$. 
% In our example, we construct the state of the network with only the edges of car mode, as the number of travelers on the edges of train mode and walk will not affect the costs of travelling on the respective edges. The state of the network $s_t$ at time $t$ is shown in the Figure \ref{fig: Example no of travelers}

% \begin{figure}[h]
% \centering
% \includegraphics[width=0.4\textwidth]{exampleNoOfTravelers.eps}
% \caption{Number of Travelers on the Network}
% \label{fig: Example no of travelers}
% \end{figure}

% \section{System's Decision Model}
Let there be a central entity (\emph{a.k.a.} the system), which evaluates the network state in terms of the overall traffic congestion and carbon emissions using a weighted multi-attribute cost. Assuming that there are $K$ attributes, each edge $e \in \mathcal{E}$ has a multi-attribute cost vector $\boldsymbol{x}(c_{e,t}) = [x_1(c_{e,t}), \ \cdots, \ x_K(c_{e,t})$]. The system evaluates the cost of each edge $e$ at time $t$ as
\begin{equation}
y(c_{e,t}) = \displaystyle \sum_{k = 1}^K a_k \cdot x_k(c_{e,t}).
\end{equation}
Since centralized systems typically have access to sensing infrastructure across the network to measure the network state in real-time, 
% consider a system denoted as $LoRI$ that designs strategic information regarding the state of the transportation network and presents this information as a signal to the travelers on the network in order to steer the travelers' decisions towards socially optimal outcomes. The system's motive is to reduce congestion and CO emission across the entire network. Consequently, the system's cost is designed based on the congestion and CO emission across the entire network. System designs and presents strategic information that minimizes it's own cost functions i.e., minimize the overall congestion and CO emission on the transportation network. As the system's cost function depend on the entire network, 
we assume that the system has greater information regarding the current state $s_t$ than the travelers.

In this paper, we assume that the system constructs a multi-dimensional signal 
% that reveals the state transition probability of the network to travelers in order to steer their decision towards socially optimal route choices
% The signaling policy adopted by the system to steer $\ell^{th}$ traveler's decision is defined as 
$\boldsymbol{\mu}_{\ell, t} = \left[ \mu_{\ell, e, t} \right]_{e \in \mathcal{E}}$ to steer $\ell^{th}$ traveler's decision, where 
\begin{equation}
\mu_{\ell, e,t}(\eta, \lambda) = [\mathbb{P}_{\ell}(c_{e,t+1} = \lambda | c_{e,t} = \eta)]_{\lambda = 0}^{c_e},
\label{eqn: System signalling policy}
\end{equation}
is the state transition probability shared by the system to the $\ell^{th}$ traveler. The system constructs this signal with the goal of steering travelers' decisions towards system's optimal (a.k.a. social welfare). 
% \textcolor{red}{SD: has a lot of concepts but would they be perceived as abstract? An illustrative diagram showing state transitions and the signal construction, for the example input network $\mathcal{G}$, would help}

Note that the overall system cost after a finite time horizon $T$ depends on decisions taken by all the active travelers and all the signals presented to the active travelers. It comprises of both past and future costs, and is given by
\begin{equation}
\begin{array}{lcl}
U_{0,T} \left( \boldsymbol{\mu}_T, \boldsymbol{p}_T \right) & = & \displaystyle 
% \sum_{t = 1}^T \sum_{e \in \mathcal{E}} y(c_{e,t}) + \sum_{t = T+1}^\infty 
\sum_{t = 1}^T \sum_{e \in \mathcal{E}} \sum_{\lambda = 1}^{\infty} \psi_{e,t}(\lambda) y(\lambda),
\end{array}
\end{equation}
where $\boldsymbol{\mu}_T = [\boldsymbol{\mu}_{1,T}, \cdots, \boldsymbol{\mu}_{\Lambda_T, T}]$ is the signal profile sent to all the travelers in the network; $\boldsymbol{p}_T = [p_{1, T}, \cdots, p_{\Lambda_T, T}]$ is the path profile chosen by the travelers; and $\psi_{e,t}(\lambda)$ denotes the \emph{a priori} system's belief probability regarding the state of edge $e$ being $c_{e,t} = \lambda$ at time $t$. 
% \textcolor{red}{SD: clarify - doesn't sound good}
Then, we define the system's rationality as follows: 
\begin{defn}
The system's motive is to minimize its cost function that depends on all the travelers' decisions and the signals presented by the system. The motive is given by:
\begin{equation}
\displaystyle \min_{\boldsymbol{\mu}_T} \ \displaystyle U_{0,T} \left( \boldsymbol{\mu}_T, \boldsymbol{p}_T \right)
\end{equation}
\label{Defn: System Rationality}
\end{defn}

\vspace{-0.2in}

\begin{figure}[!t]
\centering
\includegraphics[width=0.4\textwidth]{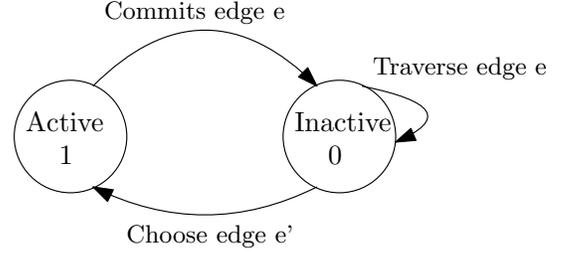}
\caption{State Transitions at the $\ell^{th}$ Traveler}
\label{fig:flow}
\end{figure}

% \section{Traveler's Decision Model}
% Let us consider that there are $\{1, 2, \cdots, \Lambda_t\}$ number of travelers at time $t$ on the transportation network. 
Although these signals can be revealed by the system at any time, the travelers can take advantage of this information and change their path only when they are present at some node. We label such agents as \emph{active} travelers. In other words, we can define the state of the $\ell^{th}$ traveler at time $t$ as
\begin{equation}
\alpha_{\ell,t} = 
\begin{cases}
1 & \text{if the } \ell^{th} \text{ traveler is active,}
\\[2ex]
0 & \text{if the } \ell^{th} \text{ traveler is inactive.}
\end{cases}
\end{equation}
In other words, an active traveler's state gets updated to an inactive state as soon as an active traveler chooses the next edge, and remains so until he/she traverses that edge completely and reaches the other vertex as shown in Figure \ref{fig:flow}. That is, $c_{e,t}$ is equal to the total number of inactive travelers $\{\alpha_{\ell,t} = 0\}$ on edge $e$ at time $t$. 
% \begin{defn}
% The number of travelers on edge $e \in \mathcal{E}$ at time $t$ is given by 
% \begin{equation}
% c_{e,t} = \sum_{\ell \in \mathcal{L}_t}\left(1-\alpha_{\ell, t} \right) \end{equation}
% \end{defn}
% Therefore, the total number of travelers on the network at time $t$ is given as: 
% \begin{equation}
%     \Lambda_t = \sum_{e \in \mathcal{E}}c_{e,t} + \sum_{\ell \in \mathcal{L}_t}\left(\alpha_{\ell, t} \right)+ \zeta
% \end{equation}
% where  $\zeta$ captures the number of travelers arriving onto the network and the number of travelers departing the network at time $t$.

Furthermore, we assume that the travelers cannot fully observe the true network state $s_t$ at any given time, but can construct a multi-dimensional belief $\phi_{\ell, t} = \left[ \phi_{\ell,e,t} \right]_{e \in \mathcal{E}}$ about $s_t$ at time $t$ based on prior experiences, where  
\begin{equation}
\phi_{\ell,e,t} = \Big\{ \phi_{\ell, e,t}(c) \Big\}_{c = 0}^{\infty}
\label{Eqn: Traveler Prior Belief}
\end{equation}
is the traveler's belief vector regarding the state of the edge $e \in \mathcal{E}$ at time $t$, and $\phi_{\ell, e,t}(c) = \mathbb{P}_{\ell}(c_{e,t} = c)$.
% For example, the prior belief $\phi_{\ell, t}$ of first traveller at time $t$ for $\lambda = \{0,1,2,3,4,5,6,7\}$ and $5$ edges in our example model is given as: 
% \begin{equation}
% \phi_{1, t} = 
% \begin{bmatrix}
% 0.70223058 &0.33170257 &0.44120985 &0.06657635 &0.27147509\\
% 0.76594134 &0.71349755 &0.19539137 &0.67759503 &0.61353863\\
% 0.83010137 &0.63931286 &0.07829764 &0.65919408 &0.86043455\\
% 0.29353045 &0.73784562 &0.79626031 &0.09300651 &0.54702768\\
% 0.01405735 &0.76015463 &0.20018205 &0.60737338 &0.88399725\\
% 0.09321457 &0.85998563 &0.38081619 &0.15700366 &0.09886448\\
% 0.14941407 &0.91636609 &0.51331971 &0.32710144 &0.47562184
% \end{bmatrix}
% \end{equation}
% Let us assume that there are $K$ attributes regarding which every edge $e \in \mathcal{E}$ is evaluated. Let $\boldsymbol{x}(c_{e,t}) = x_1(c_{e,t}), \ \cdots, \ x_K(c_{e,t})$. 
Assuming that the $\ell^{th}$ traveler's multi-attribute cost\footnote{If some attribute $k$ is not applicable to a given edge $e \in E$, then we let $x_k(c_e) = 0$. For example, the attribute `CO emissions' is not applicable to all the edges of mode "walking", for these edges, we let $x_{CO}(c_e) = 0$. } on edge $e$ at time $t$ is a weighted linear combination of all attribute-wise edge costs $\boldsymbol{x}(c_e,t)$, as given by 
% \textcolor{red}{SD: getting notation heavy! Are intuitions clear?}
\vspace{-0.1in}
\begin{equation}
z_{\ell}(c_{e,t}) = \displaystyle \sum_{k = 1}^K b_{\ell, k} \cdot x_k(c_e,t),
\end{equation}
we model the $\ell^{th}$ traveler's stochastic expected cost for choosing a path $p_{\ell,T}$ as 
\begin{equation}
V_{\ell,T}(\boldsymbol{\mu}_{\ell,T}, \pi_{\ell,T}) = %U_{\ell,T}(\mu_{\ell,T}, p_{\ell,T}) + 
\mathbb{E}_{\pi_{\ell, T}} \left[ U_{\ell,T} \left( \mu_{\ell,T}, p_{\ell, T} \right)\right] + \epsilon_{p_{\ell,T}},
\label{Eqn: random expected cost}
\end{equation}
where $\pi_{\ell,T}$ is a probability distribution over the set of all paths $\mathcal{P}_{\ell}$, $U_{\ell,T} \left( \boldsymbol{\mu}_{\ell,T}, \boldsymbol{p}_{\ell, T} \right)$ denotes the nominal (known) expected cost of the traveler, and $\epsilon_{\boldsymbol{p}_{\ell,T}}$ is the noise (random parameter) term that captures any uncertainty regarding $\ell^{th}$ traveler's rationality.
% \begin{assumption}
% System knows $\boldsymbol{b} = \{ b_{\ell,1}, \cdots, b_{\ell,K} \}$, i.e. $\ell^{th}$ traveler's cost, for all $\ell \in \mathcal{L}$.
% \end{assumption}
% A traveler will choose a path depending on his/her costs.
The decision policy adopted by the $\ell^{th}$ traveler at time $t$ is denoted as the path $\boldsymbol{p}_{\ell,t} 
% = \{ e_{\ell,0}, \cdots, e_{\ell,L} \} 
\in \mathcal{P}_{\ell}$, where $\mathcal{P}_{\ell}$ represents the set of all paths available for the $l^{th}$ traveler. 
% In practice, it is quite difficult to identify all the attributes that a traveler might be interested, in choosing his/her paths. Furthermore, there have been many papers in psychology literature that demonstrates how people deviate consistently from expected utility theory. In addition, it is practically not possible for a traveler to know the entire network state $s_t$ at every time $t$. Therefore, in this paper, we assume that the travelers minimize random costs, as opposed to deterministic expected costs. Formally, we model the $\ell^{th}$ traveler's cost as 
% \begin{equation}
% V_{\ell,T}(\mu_{\ell,T}, p_{\ell,T}) = %U_{\ell,T}(\mu_{\ell,T}, p_{\ell,T}) + 
% U_{\ell,T} \left( \mu_{\ell,T}, p_{\ell, T} \right) + \epsilon_{p_{\ell,T}},
% \label{Eqn: random expected cost}
% \end{equation}
% where $\mu_{\ell,T}$ is the signal given by the system; $U_{\ell,T} \left( \mu_{\ell,T}, p_{\ell, T} \right)$ denoted the known cost the traveler attains; and $\epsilon_{p_{\ell,T}}$ is the noise (random parameter) denoting any uncertainty regarding $\ell^{th}$ traveler's model at time $T$.

Let $\boldsymbol{p}_{\ell, 1:T}$ denote the sequence of edges that the $\ell^{th}$ traveler has already taken (committed) until time $T$. Then, the $\ell^{th}$ traveler's expected cost $U_{\ell, T}\left( \boldsymbol{\mu}_{\ell,T}, \boldsymbol{p}_{\ell, T} \right)$ comprises of two terms: the incurred (deterministic) cost from traversed, and the future (unknown) cost from the remaining path to be traversed. In other words, we have
\begin{equation}
\begin{array}{l}
U_{\ell,T} \left( \boldsymbol{\mu}_{\ell,T}, \boldsymbol{p}_{\ell, T} \right) = \displaystyle \sum_{e \in \boldsymbol{p}_{\ell, 1:T}} z_{\ell}(c_{e,t_{\ell,e}})
\\[2ex]
\qquad + \displaystyle \sum_{e \in \boldsymbol{p}_{\ell,T}-\boldsymbol{p}_{\ell,1:T}} \left(\displaystyle \sum_{\lambda = 1}^{\infty} \phi_{\ell,e,t}(\lambda) \cdot z_{\ell}(\lambda) \right),
\end{array}
\end{equation}
where $t_{\ell,e}$ is the time at which the traveler is at the head of edge $e$, and $\boldsymbol{p}_{\ell,T} - \boldsymbol{p}_{\ell,1:T}$ represents the sequence of edges that the traveler will travel in the future, if he/she continues to stay on the same decision policy $\boldsymbol{p}_{\ell,T}$. Then, the traveler's rationality is defined as follows: 
\begin{defn}
The traveler's motive is to minimize the random cost function that depends on the signals presented by the system and the path chosen by the traveler, which is given as:
\begin{equation}
\displaystyle \min_{\pi \in \Delta(\mathcal{P}_{\ell})} \ \displaystyle V_{\ell,T} \left( \boldsymbol{\mu}_{\ell,T}, \pi_{\ell, T} \right),
\end{equation}
\label{Defn: Traveler Rationality}
\end{defn}

% \begin{align*}
%     a_1, b_1, c_1, d_1& \rightarrow 3.6955824319468253, \\
%     a_1, b_1, d_1 & \rightarrow 4.468814930946907, \\
%     a_1, b_1, b_3, c_3, c_1, d_1 & \rightarrow 6.9686485142471035, \\
%     a_1, c_1, d_1 & \rightarrow 3.7108106428088807, \\
%     a_1, b_1, b_3, c_3, c_2, a_2, b_2, d_2 & \rightarrow  13.734324257123552, \\
%     a_2, b_2, b_1, c_1, d_1 & \rightarrow 5.461258174823273, \\
%     a_2, b_2, b_2, d_1 & \rightarrow 6.2344906738233545, \\
%     a_2, b_2, b_1, b_3, c_3, c_1, d_1 & \rightarrow  8.734324257123552, \\
%     a_2, b_2, b_3, c_3, c_1, d_1 & \rightarrow  7.734324257123552, \\
%     a_2, b_2, d_2 & \rightarrow 5.25
% \end{align*}

% \section{Problem Formulation}
Given that both the system and travelers have non-identical utilities (i.e., mismatched motives), it is natural to model their interaction as a one-shot Stackelberg-Quantal-Response (SQR) game, where the system commits to its signaling strategy as defined in Definition \ref{Defn: System Rationality}, before travelers choose their stochastic policies as per Definition \ref{Defn: Traveler Rationality} \cite{fudenberg1991game}. 
\begin{defn}
The equilibrium of an SQR game between the system and travellers is defined as the pair $(\mu^*_{\ell, t}, \pi^*_{\ell,t})$, where
\begin{equation}
\begin{array}{rl}
\mu^*_{\ell, t} \triangleq \displaystyle \argmin_{\mu_{\ell, t}}  & \displaystyle U_{0,T} \left( \mu_{\ell,t}, \mu_{-\ell,t}, p_{\ell,t}, p_{-\ell,t} \right)
\\[2ex]
\pi^*_{\ell,t} \triangleq \displaystyle \argmin_{\pi_{\ell,t}} & \displaystyle V_{\ell,t}\left(\mu^*_{\ell,t}, \pi_{\ell,t}\right)
% \\[2ex]
% \text{where} & \pi_{\ell,T}(p_{\ell,T}) = \displaystyle \frac{\exp{\left( \alpha \cdot U_{\ell,T}(\mu_{\ell,T}, p_{\ell,T}) \right)}}{ \displaystyle \sum_{p_{\ell,T}' \in \mathcal{P}_{\ell,T}} \exp{\left( \alpha \cdot U_{\ell,T}(\mu_{\ell,T}, p_{\ell,T}') \right)}}
\end{array}
\label{Problem: SQR Game}
\end{equation}
\label{Defn: SQR Equilibrium}
\vspace{-1ex}
\end{defn}

% However, since travelers exhibit a random cost minimization rationality and each traveler's decisions depend on all the other active travelers' decisions (Quantal Response Equilibrium), we model the interaction between the system and travelers as a bi-level \textit{"Stackelberg-Quantal Reponse Game" (SQR)}.

Similar to solving traditional Stackelberg-Nash games, we propose a novel solution approach named \emph{LoRI} based on backward induction, which evaluates travelers' quantal response equilibrium as a function of system's signal $\boldsymbol{\mu}_T$, and then evaluate the best response signal at the system. We present the technical details of our approach in the following section, and later analyze its performance in simulation experiments.

% Solving such games implies finding the Stackelberg Equilibrium – between  leader and followers – and also the Quantal Response equilibrium – in the lower hierarchical  level, among followers. At the first level of $SQR$, ($LoRI$) acts as a leader and present signal $\mu$ to travelers at time $t$. At the second level, the followers i.e., the travelers observes the system's signal $\mu$ and play a normal form game and choose a path $p \in \mathcal{P}$ based on the quantal choice model. 

\section{Equilibrium Analysis}

\begin{algorithm}[!t]
\caption{Network State Transition}
\KwData{Network State $s_t$, Current time $t$, Traveler path $p_{\ell,t}$}
\KwResult{Network State}
\For{$e \in p_{\ell,t}$}
    {
    $d[e] \gets t + TT_e(s_t[e])$ \;
    }
$keyList \gets [*d.keys()]$\;
$n[keyList[0]] \gets n[keyList[0]]+1$\;
$time[t] \gets n$\;
\For{$i$ in $keyList$}
    {
    $n[i] \gets n[i]-1$\;
    $n[i+1] \gets n[i+1]+1$\;
    $t[d[i]] \gets n$
    }
% \For{$key$ in $t$}
%     {
%     $l[key] = s_t + t[key]$ 
%     }
\label{Algo: Network State Transition}
\end{algorithm}
% \subsection{Computing Path Costs}
In order to carry out equilibrium analysis, it is necessary to evaluate the path costs at the $\ell^{th}$ traveler, which depend on thee network state. However, given that the network state evolves over time with all the active travelers' path choice updates, we first compute the network statee based on travelers' strategy profiles using Algorithm \ref{Algo: Network State Transition}. Given the network state, we evaluate the cost of traversing a path $p_{\ell,t} \in \mathcal{P}_{\ell,t}$ at the $\ell^{th}$ traveler using Algorithm \ref{Algo: Cost Matrix of traveler}. Note that the term $z_e(s_t[e])$ in Algorithm \ref{Algo: Cost Matrix of traveler} represents the cost of traveling on edge $e$ at time $t$ at the $\ell^{th}$ traveler, when its state is given by $s_t[e] = c_{e,t}$. Given the cost matrix, we now proceed to evaluating the equilibrium of the proposed SQR game using backward induction, i.e. evaluate travelers' QRE as a function of system's signal, and then compute the best-response signal at the system.

% In this paper, we evaluate the $\ell^{th}$ traveler's path costs at time $t$ depends on all the other active travelers path choices and each active traveler's path choice affects the network state. We use Algorithm \ref{Algo: Network State Transition} to find the network state based on $\ell^{th}$ traveler's chosen path $p_{\ell, t}$. 

% %\vspace{-0.1in}
% The cost of each available path $p_{\ell, t} \in \mathcal{P}_{\ell,t}$ at the $\ell^{th}$ traveler depends on the strategy profile, i.e., the paths chosen by all the active travelers at time $t$. We find the network state based on the strategy profiles using the Algorithm \ref{Algo: Network State Transition}

% The cost of $\ell^{th}$ traveler traversing a path $p_{\ell,t} \in \mathcal{P}_{\ell,t}$ at time $t$ with respect to all the possible strategy profiles of the all active travelers at time $t$, wconstructed using the Algorithm \ref{Algo: Cost Matrix of traveler}, where $z_e(s_t[e])$ is the cost of travelling on an edge $e$ at time $t$ at the $\ell^{th}$ traveler. 

\begin{algorithm}[!t]
\caption{Computing the Cost Matrix}
\KwData{Traveler $\ell$}
\KwResult{cost matrix}
\For{$p_{\ell,t} \in \mathcal{P}_{\ell,t}$}
    {
    \For{$e \in p_{\ell,t}$}
        {
        $d[e] \gets t + TT_e(c_{e,t})$ \;
        }
    \For{profile in strategy Profiles}
        {
        $ns  \gets networkStateTransition(profile)$
        \For{$key$ in d}
            {
            \For{$i$ in $ns$}
                {
                \For{$j$ in $i$}
                    {
                    \If{d[key] in range(j-10, j+10)}
                        {
                        $s_t \gets s_t + i[j]$\;
                        }
                    }
                }
            $pathCost \gets pathCost + z_e(s_t[e])$\;
            }
        $costProfile.append(pathCost)$\;
        }
    $costMatrix.append(costProfile)$\;
    }
\label{Algo: Cost Matrix of traveler}
\end{algorithm}

% \begin{algorithm}
% \caption{$z_e$}
% \KwData{Edge $e$, Traveler $\ell$, Network State $s_t$}
% \KwResult{$z_e$}
% \If{e.mode = car}
%     {
%     $cost \gets  \left(b_{\ell, TT} \right) * \left(TT_e(s_t[e])\right) + \left(b_{\ell, CO} \right) * \left(CO_e(s_t[e]) \right)$\;
%     }
% \If{e.mode = train}
%     {
%     $cost \gets  \left(b_{\ell, TT} \right) *\left(TT_e\right) + \left(b_{\ell, CO} \right) * \left(CO_e\right)$\;
%     }
% \If{e.mode = walk}
%     {
%     $cost \gets  \left(b_{\ell, TT}\right) * \left(TT_e\right) + \left(b_{\ell, CO}\right) * \left(CO_e\right)$\;
%     }
% \If{e.mode = switch}
%     {
%     $cost \gets switchCost$\;
%     }
% \label{Algo: edgeCost}
% \end{algorithm}

\subsection{Traveler's Quantal Response Analysis } \label{SubSec: QRE}

Given the system's signal $\mu_{\ell,t}$, the traveler updates his prior belief defined in Equation \eqref{Eqn: Traveler Prior Belief} using Bayes rule to obtain the following posterior belief regarding the network state:
\begin{equation}
\phi_{\ell, e,t+1}(\lambda) = \displaystyle \frac{\phi_{\ell, e,t}(\eta) \cdot \mu_{\ell, e,t}(\eta,\lambda)}{\displaystyle \sum_{\lambda = 0}^{\infty} \phi_{\ell, e,t}(\eta) \cdot \mu_{\ell, e,t}(\eta,\lambda)}.
\label{Eqn: Bayesian Update - traveler}
\end{equation}
We assume that the denominator in Equation \eqref{Eqn: Bayesian Update - traveler} always converges to some value in the region $[0,1]$ and every traveler's belief regarding the future state of the network remains stationary until the system presents a signal.

The cost that the traveler attains by choosing a path $p_{\ell,t} \in \mathcal{P}_{\ell, t}$ is decomposed into (i) known (nominal) cost at the traveler, and (ii) an unknown random cost $\epsilon_{p_{\ell,T}}$, as shown in Equation \eqref{Eqn: random expected cost}. In this paper, we assume that the noise term $\epsilon_{p_{\ell,T}}$ in the traveler's expected cost is independently, identically distributed extreme value, also known as Gumbell distribution. 
\begin{thrm}[\cite{luce1959individual}]
The $\ell^{th}$ traveler's logit choice probability for the path $p_{\ell,t}$ at time $t$ is given by:
\begin{equation}
\pi_{\ell,T}(p_{\ell,T}) = \displaystyle \frac{\exp{\left( \alpha \cdot U_{\ell,T}(\mu_{\ell,T}, p_{\ell,T}) \right)}}{ \displaystyle \sum_{p_{\ell,T}' \in \mathcal{P}_{\ell,T}} \exp{\left( \alpha \cdot U_{\ell,T}(\mu_{\ell,T}, p_{\ell,T}') \right)}},
\end{equation}
where $\alpha \geq 0$ is the parameter of the quantal response model.
\label{Thrm: Logit Probabilities}
\end{thrm}
% \begin{proof}
% The proof for Theorem \ref{Thrm: Logit Probabilities} is given in \cite{luce1959individual}
% % in Chapter 3 in the book "Discrete Choice Methods with Simulation" written by Train, Kenneth E \cite{train2009discrete}.
% \end{proof}

\vspace{-0.1in}
To compute the Quantal Response Equilibrium for the travellers, we use Gambit \cite{mckelvey2006gambit}. Gambit is a library of game theory software and tools for the construction and analysis of finite extensive and strategic games. We build a strategic game (Normal-Form game) between all the travellers and use Gambit's tool $gambit-logit$ to solve for QRE. 
Gambit computes the principle branch of the (logit) quantal response correspondence using the predictor-corrector method based on the procedure described in \cite{turocy2005dynamic}. The predictor-corrector method first generates a prediction using differential equations describing the branch of the correspondence, followed by a corrector step which refines the prediction using Newton's method for finding a zero of a function. 

% \begin{algorithm}[!t]
% \caption{QRE}
% \KwData{Traveler $\ell$, }
% \KwResult{QRE $\boldsymbol{\pi}^*$}
% $cost \gets costMatrix(\ell)$\;
% $\boldsymbol{\pi}^* \gets QRE(cost)$\;
% \label{Algo: QRE}
% \end{algorithm}

\subsection{Approximate-Response Signaling}

\begin{algorithm}[!t]
\KwData{Travelers $\Lambda_t$, Network State $s_t$}
\For{time $t = 1$ to infinity}{
    \ForAll{$\ell \in \Lambda_t$}{ 
        \If{$\alpha_{\ell,t} = 1$}{ \Comment*[r]{If traveler is active} $activeTravelers.append(\ell)$
        }
    }
    \ForAll{$\ell \in activeTravelers$}
        {
        $cost \gets costMatrix(\ell)$\;
        $\boldsymbol{\pi}^*_t \gets QRE(cost)$\;
        \While{$variable$ is a right stochastic matrix}
            {
            $\mu^* \gets \displaystyle \min \displaystyle U_{0,T} \left( \boldsymbol{\mu_T}, \boldsymbol{p_T} \right)$ \;
            }
        $chosenPath \gets path(\ell, \mu^*, s_t)$ \; 
        $e \gets chosenPath[currentEdge]$\;
        $d[\ell] \gets t + TT_{\ell, e}(c_{e,t})$\;
        }
    \ForAll{$\ell \in d$}
        {
        \If{$t > TT_{\ell, e}$}
            {
            $activeTravelers.remove(\ell)$
            }
        }
}
\caption{LoRI}
\label{Algo: LoRI}
\end{algorithm}

At any time $t$, let there be a total of $\Lambda_t$ travelers on the network. Assuming that the $\ell^{th}$ traveler is on edge $e_{\ell,t}$ at time $t$ due to the decision $p_{\ell,t}$, we can compute the total number of travelers on edge $e \in \mathcal{E}$ at the time $t$ as 
\begin{equation}
c_{e,t} = \displaystyle \sum_{\ell = 1}^{\Lambda_t} \mathds{1}(e_{\ell,t} = e),
\label{Eqn: Number of travelers}
\end{equation}
where $\mathds{1}(\cdot)$ represents the indicator function which takes the value 1 whenever the argument holds true. Given $c_{e,t}$ at time $t$ on every edge $e \in \mathcal{E}$, we can now compute the state transition probability $\psi_{e,t+1}$ as follows:
\begin{equation}
\scalemath{0.925}{
\begin{array}{l}
\psi_{e,t+1}(\eta, \lambda|\Lambda_t) =  \displaystyle \mathbb{P} \left( \left. c_{e,t+1} = \lambda \ \right| \  c_{e,t} = \eta \right)
\\[2ex]
\qquad = \displaystyle \mathbb{P} \left( \left. \sum_{\ell = 1}^{\Lambda_{t+1}} \mathds{1}(e_{\ell,t+1} = e) = \lambda \ \right| \  \sum_{\ell = 1}^{\Lambda_t} \mathds{1}(e_{\ell,t} = e) = \eta \right).
\end{array}
}
\label{Eqn: indicator equals lambda given eta}
\end{equation}
Let $\rho_{\ell, t}(e,e')$ denote the probability that the $\ell^{th}$ traveler is present on edge $e$ at time $t$ given that he is on edge $e'$ at time $t-1$. Then, the state transition probability $\psi_{e,t+1}$ can be evaluated using the following recursive relation: 
\begin{equation}
\begin{array}{l}
\psi_{e,t+1}(\eta,\lambda|\Lambda_t) = \rho_{\ell,t+1}(e,e')\cdot \psi_{e,t+1}(\eta, \lambda-1|\Lambda_t - 1)
\\[1ex]
\qquad + (1-\rho_{\ell,t+1}(e,e'))\cdot \psi_{e,t+1}(\eta, \lambda|\Lambda_t - 1)    
\end{array}
\label{Eqn: State transition recursive relation}
\end{equation}
where 
\begin{equation}
\begin{array}{l}
\rho_{\ell,t+1}(e,e',\alpha_{\ell,t}) = 
\\[2ex]
\qquad \displaystyle \sum_{p_{\ell,t} \in \mathcal{P}_{l,t}}\pi_{\ell,t}(p_{\ell,t}|\mu_{\ell,t}, e \in p_{\ell,t}, e' \in p_{\ell,t-1}, \alpha_{\ell,t} = 1)
\end{array}
\label{Eqn: Rho in terms of Pi}
\end{equation}

The leader's optimal strategy is to minimize its cost $U_{0,T}$ which can be computed as: 
\begin{equation}
\begin{array}{rl}
\displaystyle \min_{\mu_{\ell,T}} & \displaystyle U_{0,T} \left( \mu_{\ell}, \mu_{-\ell}, \pi_{\ell, T}(p_{\ell, T}|\mu_{\ell,T}), p_{-\ell, T} \right) 
% \\[3ex]
%\text{subject to} & \text{1. } \displaystyle U_{\ell,T} \left( \mu_{\ell,T}, \pi_{\ell, T}(p_{\ell, T}|\mu_{\ell,T}) \right) \leq U_{\ell,T} \left( \mu_{\ell,T}, \pi_{\ell, T}(p'_{\ell, T}) \right), 
%\\[2ex]
%& \text{2. } \displaystyle \sum_{p \in \mathcal{P_{\ell, T}}} \pi_{\ell, T}(p) = 1, \text{ and } \pi_{\ell, T}(p) \geq 0.
\end{array}
\tag{P1}
\end{equation}
% \begin{equation}
% \begin{array}{rl}
% \displaystyle \minimize_{\mu_{\ell,e,T}} & \displaystyle \sum_{t = 1}^T \sum_{e \in \mathcal{E}} y(c_{e,t}) + \sum_{t = T+1}^\infty \sum_{e \in \mathcal{E}} \sum_{\lambda = 1}^{\infty} \psi_{e,t}(\lambda) y(\lambda)
% \\[3ex]
% %\text{subject to} & \text{1. } \displaystyle U_{\ell,T} \left( \mu_{\ell,T}, \pi_{\ell, T}(p_{\ell, T}|\mu_{\ell,T}) \right) \leq U_{\ell,T} \left( \mu_{\ell,T}, \pi_{\ell, T}(p'_{\ell, T}) \right), 
% %\\[2ex]
% %& \text{2. } \displaystyle \sum_{p \in \mathcal{P_{\ell, T}}} \pi_{\ell, T}(p) = 1, \text{ and } \pi_{\ell, T}(p) \geq 0.
% \end{array}
% \end{equation}
Using Equation \eqref{Eqn: State transition recursive relation}, we write the term $\psi_{e,t}(\lambda)$ and expand $U_{0,T}$ as shown in Equation \eqref{Eqn: expanded system cost}.
\begin{figure*}[!t]
\begin{equation}
\begin{array}{lcl}
U_{0,T} & = & \displaystyle \sum_{t = 1}^T \sum_{e \in \mathcal{E}} y(c_{e,t}) + \sum_{t = T+1}^\infty \sum_{e \in \mathcal{E}} \sum_{\lambda = 1}^{\infty} \Big[ \rho_{\ell,t}(e,e')\cdot \psi_{e,t}(\eta, \lambda-1|\Lambda_{t-1} - 1)
\\[0.5ex]
& & \qquad \qquad \qquad \qquad \qquad \qquad \qquad \qquad  
+ \Big( 1-\rho_{\ell,t}(e,e') \Big) \cdot \psi_{e,t}(\eta, \lambda|\Lambda_{t-1} - 1) y(\lambda) \Big]
\\[2ex]
& = & \displaystyle \sum_{t = 1}^T \sum_{e \in \mathcal{E}} y(c_{e,t}) + \sum_{t = T+1}^\infty \sum_{e \in \mathcal{E}} \sum_{\lambda = 1}^{\infty} \left[ \left( \sum_{p_{\ell,t} \in \mathcal{P}_{l,t}}\pi_{\ell,t}(p_{\ell,t}|\mu_{\ell,t}) \right) \cdot \psi_{e,t}(\eta, \lambda-1|\Lambda_{t-1} - 1) \right.
\\[1.5ex]
& & \qquad \qquad \qquad \qquad \qquad \qquad \qquad \qquad \quad \left. + \left( 1- \displaystyle \sum_{p_{\ell,t} \in \mathcal{P}_{l,t}}\pi_{\ell,t}(p_{\ell,t}|\mu_{\ell,t}) \right) \cdot \psi_{e,t}(\eta, \lambda|\Lambda_{t-1} - 1) y(\lambda) \right]
% \\[3ex]
%\text{subject to} & \text{1. } \displaystyle U_{\ell,T} \left( \mu_{\ell,T}, \pi_{\ell, T}(p_{\ell, T}|\mu_{\ell,T}) \right) \leq U_{\ell,T} \left( \mu_{\ell,T}, \pi_{\ell, T}(p'_{\ell, T}) \right), 
%\\[2ex]
%& \text{2. } \displaystyle \sum_{p \in \mathcal{P_{\ell, T}}} \pi_{\ell, T}(p) = 1, \text{ and } \pi_{\ell, T}(p) \geq 0.
\end{array}
\label{Eqn: expanded system cost}
\end{equation}
\vspace{-4ex}
\end{figure*}

We further expand this using Equation \eqref{Eqn: Rho in terms of Pi}. For better representation, we write $\rho_{\ell,t+1}(e,e',\alpha_{\ell,t}) = \displaystyle \sum_{p_{\ell,t} \in \mathcal{P}_{l,t}}\pi_{\ell,t}(p_{\ell,t}|\mu_{\ell,t})$.
% \begin{figure*}[!t]
% \begin{equation}
% \begin{array}{rl}
% \displaystyle \minimize_{\mu_{\ell,T}} & 
% %\text{subject to} & \text{1. } \displaystyle U_{\ell,T} \left( \mu_{\ell,T}, \pi_{\ell, T}(p_{\ell, T}|\mu_{\ell,T}) \right) \leq U_{\ell,T} \left( \mu_{\ell,T}, \pi_{\ell, T}(p'_{\ell, T}) \right), 
% %\\[2ex]
% %& \text{2. } \displaystyle \sum_{p \in \mathcal{P_{\ell, T}}} \pi_{\ell, T}(p) = 1, \text{ and } \pi_{\ell, T}(p) \geq 0.
% \end{array}
% \end{equation}
% \end{figure*}
By Definition \ref{eqn: System signalling policy}, $\mu_{\ell,t} = [\mu_{\ell, e, t}]_{e \in \mathcal{E}}$ is a vector of $\mu_{\ell, e, t}$ for every edge $e \in \mathcal{E}$ in the network. $\mu_{\ell, e, t} = [\mathbb{P}_{\ell}(c_{e,t+1} = \lambda | c_{e,t} = \eta)]_{\lambda = 0}^{c_e}$ is a vector of probabilities $\mathbb{P}_{\ell}(c_{e,t+1} = \lambda | c_{e,t} = \eta)$ for all possible values of $\lambda$. We assume that the upper bound of $\lambda$ is the capacity $c_e$ of the edge $e$. 
% \begin{equation}
% \mu_{\ell,e,t} = 
% \begin{bmatrix}
% \mathbb{P}_{\ell}(c_{e,t+1} = 0 | c_{e,t} = \eta)&\cdots&\mathbb{P}_{\ell}(c_{e,t+1} = c_e | c_{e,t} = \eta)
% \end{bmatrix},
% \end{equation}
% $\mu_{\ell,t}$ is a vector comprising of $\mu_{\ell,e,t}$ vectors for all $e \in \mathcal{E}$
% \begin{equation}
% \mu_{\ell,t} = 
% \begin{bmatrix}
% [\mu_{\ell,1,t}] & [\mu_{\ell,2,t}] & [\mu_{\ell,3,t}] & \cdots & [\mu_{\ell,\mathcal{E},t}]
% \end{bmatrix}    
% \end{equation}
% We denote the $\mu_{\ell,t}$ search space as $\Omega_{\mu_{\ell,t}}$. To reduce $\Omega_{\mu_{\ell,t}}$, we consider only discrete probabilities $\mathbb{P}_{\ell}(c_{e,t+1} = \lambda | c_{e,t} = \eta) \in \{0, 0.25, 0.5, 0.75, 1\}$. Consequently, for every edge $e$, there are $5^{c_e}$ number of possible $\mu_{\ell,e,t}$ vectors. The search space $\Omega_{\mu_{\ell,t}}$ comprises of all possible different $\mu_{\ell,t}$s constructed using $5^{c_e}$ different possible $\mu_{\ell,e,t}$ vectors for all edges $e \in \mathcal{E}$.
Since the system has a cost minimization rationality, it will send a signal $\mu_{\ell,t}$ at time $t$ to the $\ell^{th}$ traveler such that the path chosen by the traveler minimizes the system's overall cost. 
% Upon receiving the signal the $\ell^{th}$ traveler updates his/her belief $\phi_{\ell,t}$ regarding the state of the network. The traveler will update his/her beliefs as shown in Equation \eqref{Eqn: Bayesian Update - traveler}.  
Since we have a leader-follower game, we use backward induction to solve for the optimal leader strategy, i.e., the optimal signal at the  system. System cannot send signal to every traveller at the same time as every traveller's decision depend on all the other travellers as well. Therefore, for every time step, the system sends to signals travellers in a round-robin fashion. The system sends a signal $\mu_{\ell,t}$ to the $\ell^{th}$ traveler while all the other travelers are fixed on their respective paths.
% \centering
% \includegraphics[width=\textwidth]{algorithmFlow.eps}
% \caption{Algorithm Flow}
% \label{fig:flow}
% \end{figure*}

% \begin{algorithm}

% \For{$\ell \in \{1,2,\cdots,\Lambda_t\}$} 
%     {
%     \For{$\mu_{\ell, t} \in \Omega_{\ell,t}$} 
%         {    
%         $\phi_{\ell, t+1} \gets \phi_{\ell, t}(\mu_{\ell, t})$ 
%         \Comment*[r]{Updates traveler's beliefs}
%          $\boldsymbol{\pi}_{t}(\boldsymbol{p}_{t}) \gets QRE[\Lambda_t]$ 
%         \Comment*[r]{QRE gives $\ell^{th}$ traveler's probability of choosing path $p_{\ell,t}$}
%         $systemCost.add(U_{0}(\mu_{\ell,t},\pi_{\ell,t}(p_{\ell,t})))$ 
%         \Comment*[r]{Append system's cost of signal $\mu_{\ell,t}$ to an array}
%         }
%     {$\mu_{\ell,t}^* \gets min(systemCost[])$} 
%     \Comment*[r]{Optimal signal yields the minimum cost at the system} 
%     $\mu^*.add(\mu_{\ell,t}^*$) \Comment*[r]{Array of optimal signals for all traveler}
% } 
% \end{algorithm}

% \begin{algorithm}[!t]
% \caption{Approximate Signal}
% \KwData{Traveler $\ell_{t}$, }
% \KwResult{approximate Signal $\mu^*$}
% $systemCost \gets sysCost(s_t)$\;
% $variable \gets$ a matrix of size $numEdges \times maxTravelers$\;
% \While{$variable$ is a right stochastic matrix}
%     {
%     $\mu^* \gets \displaystyle \min_{\mu_{\ell,T}} \displaystyle U_{0,T} \left( \mu_{\ell}, \mu_{-\ell}, \pi_{\ell, T}(p_{\ell, T}|\mu_{\ell,T}), p_{-\ell, T} \right)  \ $
%     }
% \label{Algo: approximate Signaling}
% \end{algorithm}

The search space in this optimization problem comprises of all right stochastic matrices which can be shown as a convex set. However, it is analytically hard to verify whether or not, the objective function $U_{0,T}$ stated in Equation \eqref{Eqn: expanded system cost} is convex in $\mu$. Note that the term $\pi_{\ell, t}$ represents logit probabilities which are known to be non-convex. Equation \eqref{Eqn: expanded system cost} comprises of convex combination of sum of logit probabilities whose convexity properties are hard to verify. Therefore, we employ interior point algorithms to compute the approximate signal. In our simulation experiments, we use CVXPY \cite{diamond2016cvxpy} package to implement interior point search in Algorithm \ref{Algo: LoRI}. 
% Using the Algorithm \ref{Algo: approximate Signaling}, we get the probabilities $\{\pi_{1,t}(p_{1,t}), \pi_{2,t}(p_{2,t}), \cdots,  \pi_{1,t}(p_{1,t})\}$ for all the travelers. Using these probabilities, we find the state transition probabilities $\psi_{e,t}$ as shown in Equation \eqref{Eqn: State transition recursive relation}. 

% \newpage
% \begin{figure}[h]
%     \centering
%     \includegraphics[width=0.8\textwidth]{basicFlow.eps}
%     \caption{Flow Chart}
%     \label{fig:flow}
% \end{figure}

% The signal $\mu$ depends on various other functions and has the following Bayesian Network: 
% \begin{equation*}
%     \mu \rightarrow \phi \rightarrow \pi \rightarrow \psi
% \end{equation*}
\section{Results and Discussions}
In this section, we discuss our simulation experiments along with our findings in terms of the performance of LoRI, in comparison to single-source shortest path (SSSP) algorithms used by traditional navigation systems in the context of a Wheatstone network shown in Figure \ref{fig: Example Expanded Network}. We assume that SSSP algorithms are constructed based on a single attribute, namely \emph{Travel Time}, whereas our proposed algorithm (LoRI) relies on two attributes, namely \emph{Travel Time}, and \emph{CO Emissions}. Depending on the transport mode, we employed well-known cost models found in the literature, to carry out our simulation experiments. For example, travel time $TT_e$ on edge $e$ can be calculated for transport modes serviced on a road network (e.g. car, taxi, bus) using Bureau of Public Roads (BPR) formula \cite{manualbureau}: 
\begin{equation}
    TT_e(c_{e,t}) = f_e\left[1 
    + \alpha \left(\frac{n_{e,t+1}}{c_{e}}\right)^{\beta}\right],
\label{Eqn: BPR function}
\end{equation}
where $n_{e,t}$ is the number of vehicles at time $t$, $c_e$ is the capacity,  of the edge on edge $e$, $f_e$ denotes the free-flow travel time of edge $e$. $\alpha$ and $\beta$ ares constants in the BPR function (usually $\alpha$ is 0.15 and $\beta$ is 4). 
% Upon keen observation, the reader may find that the travel time will stay at the free flow travel time until the flow is very close to road capacity, after which it increases rapidly as the volume of vehicles approach the capacity.
Similarly, the rate of carbon emissions per vehicle can be calculated using a a non-linear, static emission model for network links proposed by Wallace et al. \cite{wallace1998transyt}, as shown below:
\begin{equation}
CO_{e}(TT_e(n_{e,t})) = 0.2038 TT_{e}(n_{e,t})\exp\frac{0.7962l_e}{TT_{e}(n_{e,t})}
\label{Eqn: CO emission function }
\end{equation}
where $l_e$ is the link length (in kilometers), $T_e(n_{e,t})$ is the travel time (in minutes) for link $e$, and $CO_{e}$ is measured in grams per vehicle per hour. 

The travel time for edges that support \emph{subway} mode can be extracted from their arrival and departures time. In our example network in Figure \ref{fig: Example Expanded Network}, we asumme travel times as $\{a \rightarrow b : 3, b \rightarrow d: 4, d \rightarrow c:2, c \rightarrow a: 4\}$. For simplicity, we assume CO emissions per traveler on a subway to be $0.5*travel\_time$. For the edge corresponding to \emph{walking} modality, we assume the travel time $\{b \rightarrow c: 4\}$ and CO emissions to be simply $0$. In this example, we consider the total cost of traveling on a switch edge to be $1$. 
We implement our simulation experiments in two different scenarios using the following Python packages: python-igraph v$0.9.6$, gambit v$16.0.1$, cvxpy v$1.1.14$, numpy v$1.21.1$, matplotlib v$3.4.3$ and all their dependencies.

\begin{figure}[!t]
\centering
\includegraphics[width=0.46\textwidth]{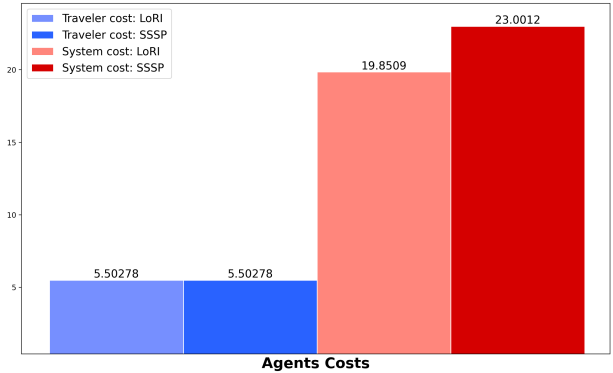}
\caption{Comparison of agents costs due to LoRI and SSSP in the first experiment under Scenario 1}
\label{fig: Comparing agents costs}
\vspace{-2ex}
\end{figure}

\subsection{Scenario 1}
We simulate three different travelers with unique origin-destination pairs: $\{(a,d), (d,b), (c,b)\}$, each of whom interacts with SSSP for a route recommendation and LoRI for network state information. In our first experiment, we assume the LoRI's weight for travel time to be $0.7$, and traveler's weights for travel time to be $\{0, 0.25, 0.5, 0.75, 1\}$. We compute the empirical average costs across different traveler motives at both traveler and system, and plot them in Figure \ref{fig: Comparing agents costs}. Although the travelers' average cost remains the same for both SSSP or LoRI, the system cost reduces by about $14\%$ when the travelers interact with the LoRI in lieu of SSSP. Specifically, LoRI reduces the congestion rate by about $\mathbf{10\%}$ and CO emissions rate by $\mathbf{4\%}$ across the entire multi-modal network. 

\begin{figure}[!t]
\centering
\includegraphics[width=0.46\textwidth]{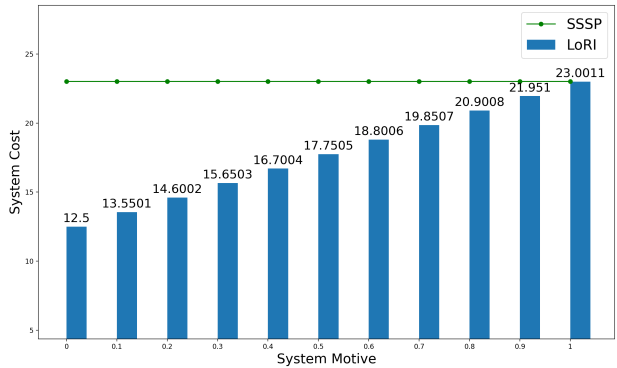}
% \vspace{-0.1in}
\caption{Comparison of system costs across different motives due to LoRI and SSSP in second experiment under Scenario 1}
\label{fig: graph 1}
% \vspace{-ex}
\end{figure}

In our second experiment, we assume that the three traveler's weights for travel time are $0.8, 0.6, 0.5$, and varied LoRI's weights across $\{0, 0.1, 0.2, \cdots, 1.0\}$. We evaluated average system costs across different origin-destination pairs and plot them as shown in Figure \ref{fig: graph 1}. It is quite evident that LoRI's costs are at least as good as that of SSSP. Specifically, system obtains a tremendous gain by adopting LoRI when there is a motive mismatch between SSSP and the system. For example, when the system's weight for travel time is $0$, the adoption of LoRI reduces the overall network congestion by $\mathbf{45\%}$. 

In our third experiment, we assume every travelers' weight for travel time to be $0.5$ for all possible origin-destinations pairs. In this experiment, we observe that LoRI successfully persuades the traveler with $\mathbf{66.66\%}$ probability, i.e. LoRI's strategically designed information was able to successfully steer travelers' routes towards socially optimal choices over $66.66\%$ of all origin-destination pairs. 

\subsection{Scenario 2}

\begin{table}[!t]
\centering
\begin{adjustbox}{width=0.45\textwidth}
\begin{tabular}{ |c|c|c| } 
 \hline
 Number of Travelers & LoRI& SSSP  \\
  \hline \hline
 1 & 20.211234567901233 & 23.624074074074073 \\ 
 \hline
 2 & 20.135617283950616 & 23.55925925925926  \\ 
 \hline
 3 &19.85057613168724 & 23.001058201058203 \\
 \hline
\end{tabular}
\end{adjustbox}
\caption{System costs with $x$ travelers interacting with LoRI under Scenario 2}
\label{Tab: percentage table}
\end{table}

\begin{table}[!t]
\centering
\begin{adjustbox}{width=0.46\textwidth}
\begin{tabular}{ |c|c|c| } 
 \hline
 Number of Travelers & LoRI& SSSP  \\
  \hline \hline
 1 & 0.24670171737670898 & 0.0011279582977294922 \\ 
 \hline
 2 & 0.41092681884765625 & 0.002007007598876953  \\ 
 \hline
 3 &2.2496159076690674 & 0.002805948257446289 \\
 \hline
 4 & 909.9852938652039 & 0.0027680397033691406 \\
 \hline
\end{tabular}
\end{adjustbox}
\caption{Run Time under Scenario 2}
\label{Tab: run time table}
\end{table}

In this scenario, we distribute 30 travelers across the entire multi-modal transportation network. 
% \textcolor{red}{How?}. 
We perform this experiment with $x$ number of travelers who interact with LoRI, while all the other travelers $(30-x)$ interact with SSSP. We assume the system's weight for travel time to be $0.7$. We calculated average system costs across different traveler paths and present them in Table \ref{Tab: percentage table}. Note that the system's social welfare consistently improves as the number of travelers interacting with LoRI increases. However, there is a significant tradeoff in terms of run time. Table \ref{Tab: run time table} shows how the runtime of LoRI and SSSP varies with $x$. Although SSSP's runtime remains almost unchanged with increasing number of travelers in our experiment, LoRI's runtime increases exponentially with increasing number of interacting travelers. This exponential increase in runtime happens because of significant increase in the number of possible signaling strategies at the system, which in turn depends on all possible combinations of all the paths available at every active travelers ($\alpha_{\ell,t} = 1$) interacting with LoRI. 
% We calculate the cost at every traveler with respect to all the possible strategy profiles at every time step. This increase in run time is the price we pay to reduce the congestion and CO emission across the entire network. 

\section{Conclusion}
In summary, we proposed a novel Stackelberg signaling framework to improve the inefficiency of selfish routing in the presence of behavioral agents. We modeled the interaction between the system and quantal response travelers as a Stackelberg game, and developed a novel approximate algorithm \emph{LoRI} that constructs strategic, personalized information regarding the state of the network. The system presents this information as a private signal to each traveler to steer their route decisions towards socially optimal outcomes. We demonstrate the performance of LoRI and compare with that of a SSSP algorithm on a Wheatstone network with multi-modal routes. We presented the tradeoff between system's costs and runtime within strategic information design framework. In the future, we will design computationally efficient, approximate algorithms at the system with better run-time performance. We will also consider strategic information design for diverse agent rationalities. 
% We will scale up LoRI to work with more realistic number of travelers on a real time multi-modal transportation network. 

\newpage
\bibliography{AAAI22.bib}

\begin{thebibliography}{29}
\providecommand{\natexlab}[1]{#1}

\bibitem[{Acemoglu et~al.(2018)Acemoglu, Makhdoumi, Malekian, and
  Ozdaglar}]{acemoglu2018informational}
Acemoglu, D.; Makhdoumi, A.; Malekian, A.; and Ozdaglar, A. 2018.
\newblock Informational Braess’ paradox: The effect of information on traffic
  congestion.
\newblock \emph{Operations Research}, 66(4): 893--917.

\bibitem[{Arnott, De~Palma, and Lindsey(1991)}]{arnott1991does}
Arnott, R.; De~Palma, A.; and Lindsey, R. 1991.
\newblock Does providing information to drivers reduce traffic congestion?
\newblock \emph{Transportation Research Part A: General}, 25(5): 309--318.

\bibitem[{Bergemann and Morris(2016)}]{bergemann2016bayes}
Bergemann, D.; and Morris, S. 2016.
\newblock Bayes correlated equilibrium and the comparison of information
  structures in games.
\newblock \emph{Theoretical Economics}, 11(2): 487--522.

\bibitem[{Bureau(2014)}]{USCensus}
Bureau, U.~C. 2014.
\newblock Biking to Work Increases 60 Percent Over Last Decade, Census Bureau
  Reports.
\newblock
  \emph{\url{https://www.census.gov/newsroom/press-releases/2014/cb14-86.html}}.

\bibitem[{Cormen et~al.(2009)Cormen, Leiserson, Rivest, and
  Stein}]{10.5555/1614191}
Cormen, T.~H.; Leiserson, C.~E.; Rivest, R.~L.; and Stein, C. 2009.
\newblock \emph{Introduction to Algorithms, Third Edition}.
\newblock The MIT Press, 3rd edition.
\newblock ISBN 0262033844.

\bibitem[{Das, Kamenica, and Mirka(2017)}]{das2017reducing}
Das, S.; Kamenica, E.; and Mirka, R. 2017.
\newblock Reducing congestion through information design.
\newblock In \emph{2017 55th annual allerton conference on communication,
  control, and computing (allerton)}, 1279--1284. IEEE.

\bibitem[{Diamond and Boyd(2016)}]{diamond2016cvxpy}
Diamond, S.; and Boyd, S. 2016.
\newblock {CVXPY}: {A} {P}ython-embedded modeling language for convex
  optimization.
\newblock \emph{Journal of Machine Learning Research}, 17(83): 1--5.

\bibitem[{Fudenberg and Tirole(1991)}]{fudenberg1991game}
Fudenberg, D.; and Tirole, J. 1991.
\newblock Game theory, 1991.
\newblock \emph{Cambridge, Massachusetts}, 393(12): 80.

\bibitem[{INRIX(2020)}]{INRIX}
INRIX. 2020.
\newblock Global Traffic Scorecard.
\newblock \emph{\url{https://inrix.com/scorecard}}.

\bibitem[{Kamenica(2019)}]{kamenica2019bayesian}
Kamenica, E. 2019.
\newblock Bayesian persuasion and information design.
\newblock \emph{Annual Review of Economics}, 11: 249--272.

\bibitem[{Lanning, Harrell, and Wang(2014)}]{lanning2014dijkstra}
Lanning, D.~R.; Harrell, G.~K.; and Wang, J. 2014.
\newblock Dijkstra's algorithm and Google maps.
\newblock In \emph{Proceedings of the 2014 ACM Southeast Regional Conference},
  1--3.

\bibitem[{Literacy(2020)}]{energy}
Literacy, E. 2020.
\newblock Energy Literacy.
\newblock \emph{\url{http://energyliteracy.com}}.

\bibitem[{Luce(1959)}]{luce1959individual}
Luce, R.~D. 1959.
\newblock Individual choice behavior, John Wiley and Sons.

\bibitem[{Mahmassani and Jayakrishnan(1991)}]{mahmassani1991system}
Mahmassani, H.~S.; and Jayakrishnan, R. 1991.
\newblock System performance and user response under real-time information in a
  congested traffic corridor.
\newblock \emph{Transportation Research Part A: General}, 25(5): 293--307.

\bibitem[{Manual(1964)}]{manualbureau}
Manual, T.~A. 1964.
\newblock Bureau of Public Roads, US Department of Commerce, 1964.
\newblock \emph{Google Scholar}.

\bibitem[{Mathevet, Perego, and Taneva(2020)}]{mathevet2020information}
Mathevet, L.; Perego, J.; and Taneva, I. 2020.
\newblock On information design in games.
\newblock \emph{Journal of Political Economy}, 128(4): 1370--1404.

\bibitem[{McKelvey, McLennan, and Turocy(2006)}]{mckelvey2006gambit}
McKelvey, R.~D.; McLennan, A.~M.; and Turocy, T.~L. 2006.
\newblock Gambit: Software tools for game theory.
\newblock \emph{\url{http://www.gambit-project.org}}.

\bibitem[{McKenzie et~al.(2015)}]{mckenzie2015drives}
McKenzie, B.; et~al. 2015.
\newblock \emph{Who Drives to Work?: Commuting by Automobile in the United
  States: 2013}.
\newblock US Department of Commerce, Economics and Statistics Administration,
  US~….

\bibitem[{Nadendla, Langbort, and Ba{\c{s}}ar(2018)}]{nadendla2018effects}
Nadendla, V. S.~S.; Langbort, C.; and Ba{\c{s}}ar, T. 2018.
\newblock Effects of subjective biases on strategic information transmission.
\newblock \emph{IEEE Transactions on Communications}, 66(12): 6040--6049.

\bibitem[{Roughgarden and Tardos(2002)}]{roughgarden2002bad}
Roughgarden, T.; and Tardos, {\'E}. 2002.
\newblock How bad is selfish routing?
\newblock \emph{Journal of the ACM (JACM)}, 49(2): 236--259.

\bibitem[{Ruggles(1949)}]{ruggles1949recent}
Ruggles, N. 1949.
\newblock Recent developments in the theory of marginal cost pricing.
\newblock \emph{The Review of Economic Studies}, 17(2): 107--126.

\bibitem[{Samal et~al.(2018)Samal, Zheng, Sun, Ratliff, and
  Dubey}]{samal2018towards}
Samal, C.; Zheng, L.; Sun, F.; Ratliff, L.~J.; and Dubey, A. 2018.
\newblock Towards a socially optimal multi-modal routing platform.
\newblock \emph{arXiv preprint arXiv:1802.10140}.

\bibitem[{Sharon et~al.(2019)Sharon, Boyles, Alkoby, and
  Stone}]{sharon2019marginal}
Sharon, G.; Boyles, S.~D.; Alkoby, S.; and Stone, P. 2019.
\newblock Marginal Cost Pricing with a Fixed Error Factor in Traffic Networks.
\newblock In \emph{AAMAS}, 1539--1546.

\bibitem[{Swamy(2007)}]{swamy2007effectiveness}
Swamy, C. 2007.
\newblock The effectiveness of Stackelberg strategies and tolls for network
  congestion games.
\newblock In \emph{SODA}, 1133--1142. Citeseer.

\bibitem[{Turocy(2005)}]{turocy2005dynamic}
Turocy, T.~L. 2005.
\newblock A dynamic homotopy interpretation of the logistic quantal response
  equilibrium correspondence.
\newblock \emph{Games and Economic Behavior}, 51(2): 243--263.

\bibitem[{Wallace et~al.(1998)Wallace, Courage, Hadi, and
  Gan}]{wallace1998transyt}
Wallace, C.; Courage, K.; Hadi, M.; and Gan, A. 1998.
\newblock TRANSYT-7F Users Guide, Methodology for Optimizing Signal Timing,
  Vol. 4.
\newblock \emph{Transportation Research Center, University of Florida,
  Gainesville}.

\bibitem[{Wash(2020)}]{INRIX2020}
Wash, K. 2020.
\newblock Congestion Costs Each American Nearly 100 hours, \$1,400 A Year.
\newblock
  \emph{\url{https://inrix.com/press-releases/2019-traffic-scorecard-us/}}.

\bibitem[{Wu and Amin(2019)}]{wu2019information}
Wu, M.; and Amin, S. 2019.
\newblock Information design for regulating traffic flows under uncertain
  network state.
\newblock In \emph{2019 57th Annual Allerton Conference on Communication,
  Control, and Computing (Allerton)}, 671--678. IEEE.

\bibitem[{Yang, Purevjav, and Li(2020)}]{yang2020marginal}
Yang, J.; Purevjav, A.-O.; and Li, S. 2020.
\newblock The marginal cost of traffic congestion and road pricing: Evidence
  from a natural experiment in Beijing.
\newblock \emph{American Economic Journal: Economic Policy}, 12(1): 418--53.

\end{thebibliography}
\end{document}